\documentclass[11pt]{llncs}
\newif\ifFull
\Fullfalse
\usepackage{t1enc}
\usepackage[utf8]{inputenc}
\usepackage{amsfonts}

\newcommand{\set}[1]{\left\{ #1\right\}}

\newcommand{\sodass}{\,:\,}
\newcommand{\setGilt}[2]{\left\{ #1\sodass #2\right\}}

\newcommand{\realrange}[2]{\left[#1, #2\right]}

\newcommand{\unitrange}[2]{\realrange{0}{1}}

\newcommand{\llabel}[1]{\label{\labelprefix:#1}}
\newcommand{\labelprefix}{} % later redefined using renewcommand

\newcommand{\discussionsize}{\small}

\newenvironment{code}{\noindent%\sf%
\begin{tabbing}%
\hspace{2em}\=\hspace{2em}\=\hspace{2em}\=\hspace{2em}\=\hspace{2em}\=%
\hspace{2em}\=\hspace{2em}\=\hspace{2em}\=\hspace{2em}\=\hspace{2em}\=%
\kill}{\end{tabbing}}

\newcommand{\labelcommand}{}
\newcommand{\captiontext}{}
\newsavebox{\codeparam}
\newcounter{lineNumber}
\newenvironment{disscodepos}[3]{%
\renewcommand{\labelcommand}{#2}%
\renewcommand{\captiontext}{#3}%
\sbox{\codeparam}{\parbox{\textwidth}{#3}}%
\begin{figure}[#1]\begin{center}\begin{code}\setcounter{lineNumber}{1}}{%
\end{code}\end{center}\caption{\llabel{\labelcommand}\captiontext}\end{figure}}

{\end{disscodepos}}

\newdimen\endofsize\endofsize=0.5em
\def\endofbeweis{~\quad\hglue\hsize minus\hsize
                 \hbox{\vrule height \endofsize width
\endofsize}\par}
\usepackage{epsfig}
\usepackage{url}
\usepackage{amsmath}
\usepackage{amssymb}
\usepackage{algorithm}
\usepackage{algorithmic}
\usepackage{pdflscape}
\usepackage{numprint}
\npdecimalsign{.} % we want . not , in numbers
\usepackage{theorem}
\usepackage{makeidx}
\usepackage[usenames,dvipsnames]{xcolor}
\usepackage{hyperref}
\usepackage{xspace}
\usepackage{wrapfig}
\usepackage{graphics}
\usepackage{todonotes}
\usepackage{booktabs}
\usepackage{multirow}
\usepackage{tabulary}
\usepackage{longtable}
\usepackage[activate={true,nocompatibility},final,tracking=true,kerning=true,spacing=true,stretch=10,shrink=30]{microtype}

\newcommand{\Id}[1]{{\texttt{#1}}}
\newcommand{\AlgName}[1]{\ensuremath{\text{{\sf #1}}}}
\newcommand{\arw}{\AlgName{ARW}}
\newcommand{\kermis}{\AlgName{KerMIS}}
\newcommand{\redumis}{\AlgName{ReduMIS}}
\newcommand{\online}{\AlgName{OnlineMIS}}

\newcommand{\etal}{et~al.~}

\usepackage{color}

\newcommand{\strash}[1]{{\color{orange}[DS: #1]}}
\newcommand{\csch}[1]{{\color{blue}[CS: #1]}}
\newcommand{\lamm}[1]{{\color{blue}[SL: #1]}}
\newcommand{\sanders}[1]{{\color{blue}[PS: #1]}}
\newcommand{\werneck}[1]{{\color{blue}[RW: #1]}}
\renewcommand{\csch}[1]{}
\renewcommand{\strash}[1]{}
\renewcommand{\lamm}[1]{}
\renewcommand{\sanders}[1]{}
\renewcommand{\werneck}[1]{}

\def\comment#1{}

\def\withcomments{
  \newcounter{mycommentcounter}
   \def\comment##1{\refstepcounter{mycommentcounter}%
    \ifhmode%
     \unskip%
     {\dimen1=\baselineskip \divide\dimen1 by 2 %
       \raise\dimen1\llap{\tiny\bfseries \textcolor{red}{-\themycommentcounter-}}}\fi%
     \marginpar[{\renewcommand{\baselinestretch}{0.8}%
       \hspace*{3em}\begin{minipage}{5em}\footnotesize [\themycommentcounter]: \raggedright ##1\end{minipage}}]{\renewcommand{\baselinestretch}{0.8}%
       \begin{minipage}{5em}\footnotesize [\themycommentcounter]: \raggedright ##1\end{minipage}}}
  }

\definecolor{darkgreen}{RGB}{0,200,100}
\definecolor{orange}{RGB}{255,80,0}

\newcommand{\Xcomment}[1]{}

\newcommand{\mytitle}{Accelerating Local Search for the \\ Maximum Independent Set Problem}

\newcommand{\detailedheader}{
\centering
\small
\setlength{\tabcolsep}{.7ex}
\begin{tabular}{lrr@{\hskip 13pt} rrr@{\hskip 13pt} rrr@{\hskip 13pt} rrr}
\toprule
\multicolumn{3}{c}{Graph}  & \multicolumn{3}{c}{\redumis{}}& \multicolumn{3}{c}{EvoMIS} & \multicolumn{3}{c}{\arw} \\
\cmidrule(r){1-3}\cmidrule(r){4-6} \cmidrule(r){7-9} \cmidrule{10-12}                             
Name                            & $n$  & Opt. & Avg.                       & Max.                       & Min.                        & Avg.                       & Max.                       & Min.              & Avg.              & Max.                       & Min. \\
\midrule                          
}

\newcommand{\detailedheadertabspeedup}{
\setlength{\tabcolsep}{.7ex}
\begin{tabulary}{\textwidth}{l@{\hskip 5pt}r@{\hskip 5pt}r@{\hskip 5pt}r@{\hskip 5pt}r@{\hskip 5pt}r}
\toprule
\multicolumn{3}{c}{Graph} & \multicolumn{3}{c}{Maximum Speedup of \online{}} \\
Name & \multicolumn{1}{c}{$n$} & \multicolumn{1}{c}{$m$} & \multicolumn{1}{c}{$s_{\arw}^{\max}$} & \multicolumn{1}{c}{$s_{\kermis}^{\max}$} & \multicolumn{1}{c}{$s_{\redumis}^{\max}$} \\ 
\cmidrule(r){1-1} \cmidrule(r){2-2} \cmidrule(r){3-3} \cmidrule(r){4-4} \cmidrule(r){5-5} \cmidrule(r){6-6}
}

\newcommand{\detailedheadertabpercent}{
\setlength{\tabcolsep}{.7ex}
\begin{tabulary}{\textwidth}{l@{\hskip 5pt}r@{\hskip 5pt}r@{\hskip 5pt}r@{\hskip 5pt}r@{\hskip 5pt}r@{\hskip 5pt}l}
\toprule
\multicolumn{1}{c}{Graph} & \multicolumn{1}{c}{\online} & \multicolumn{1}{c}{\arw} & \multicolumn{1}{c}{\kermis} & \multicolumn{1}{c}{\redumis} & \multicolumn{1}{c}{Best IS} & \multicolumn{1}{c}{Best IS}\\
Name & $t_{\mathrm{avg}}$ & $t_{\mathrm{avg}}$ & $t_{\mathrm{avg}}$ & $t_{\mathrm{avg}}$ & \multicolumn{1}{c}{Size} & \multicolumn{1}{c}{Algorithms} \\
\cmidrule(r){1-1}\cmidrule(r){2-2} \cmidrule(r){3-3} \cmidrule(r){4-4} \cmidrule(r){5-5} \cmidrule(r){6-6} \cmidrule(r){7-7}
}

\newcommand{\detailedheadertabavg}{
\setlength{\tabcolsep}{.7ex}
\begin{tabulary}{\textwidth}{l@{\hskip 5pt}r r@{\hskip 5pt}r r@{\hskip 5pt}r r@{\hskip 5pt}r r}
\toprule
\multicolumn{1}{c}{Graph} & \multicolumn{2}{c}{\online} & \multicolumn{2}{c}{\arw} & \multicolumn{2}{c}{\kermis} & \multicolumn{2}{c}{\redumis} \\
Name & \multicolumn{1}{c}{Avg.} & $t_{\mathrm{avg}}$ & \multicolumn{1}{c}{Avg.} & $t_{\mathrm{avg}}$ & \multicolumn{1}{c}{Avg.}& $t_{\mathrm{avg}}$ &\multicolumn{1}{c}{Avg.} & $t_{\mathrm{avg}}$ \\
\cmidrule(r){1-1}\cmidrule(r){2-3} \cmidrule(r){4-5}                        \cmidrule(r){6-7}\cmidrule(r){8-9}
}

\begin{document}
\title{\mytitle}
\author{Jakob Dahlum, Sebastian Lamm, Peter Sanders,\\ Christian Schulz, Darren Strash and Renato F. Werneck\\ 
	\textit{Karlsruhe Institute of Technology},
	\textit{Karlsruhe, Germany} \\
    \texttt{\{dahlum, lamm\}}\texttt{@ira.uka.de},\\ \texttt{\{sanders, christian.schulz, strash\}}\texttt{@kit.edu}, \\
	\textit{San Francisco, USA} \\
    \texttt{rwerneck}\texttt{@acm.org}}
\institute{}
\date{}

\maketitle
\begin{abstract}
Computing high-quality independent sets quickly is an important problem in combinatorial optimization.
Several recent algorithms have shown that kernelization techniques can be used to find exact maximum independent sets in medium-sized sparse graphs, as well as high-quality independent sets in huge sparse graphs that are intractable for exact (exponential-time) algorithms. However, a major drawback of these algorithms is that they require significant preprocessing overhead, and therefore cannot be used to find a high-quality independent set quickly.

In this paper, we show that performing simple kernelization techniques in an online fashion significantly boosts the performance of local search, and is much faster than pre-computing a kernel using advanced techniques. In addition, we show that cutting high-degree vertices can boost local search performance even further, especially on huge (sparse) complex networks.
Our experiments show that we can drastically speed up the computation of large independent sets compared to other state-of-the-art algorithms, while also producing results that are very close to the best known solutions. 
\end{abstract}

%\vfill
%\pagebreak
%%%%%%%%%%%%%%%%%%%%%%%%%%%%%
\section{Introduction}
\label{s:introduction}
The maximum independent set problem is a classic NP-hard problem~\cite{DBLP:books/fm/GareyJ79} with applications spanning many fields, such as classification theory, information retrieval, computer vision~\cite{feo1994greedy}, computer graphics~\cite{sander-mesh-2008}, map labeling~\cite{gemsa2014dynamiclabel} and routing in road networks~\cite{kieritz-contraction-2010}. Given a graph $G=(V,E)$, our goal is to compute a maximum cardinality set of vertices $\mathcal{I}\subseteq V$ such that no vertices in $\mathcal{I}$ are adjacent to one another. Such a set is called a \emph{maximum independent set} (MIS). %The maximum independent set problem has applications spanning many disciplines. 

\subsection{Previous Work}
\label{s:related}

Since the MIS problem is NP-hard,
all known exact algorithms for these problems take exponential time, making \emph{large} graphs infeasible to solve in practice. 
Instead, heuristic algorithms such as local search are used to efficiently compute high-quality independent sets. For many practical instances, some local search algorithms even quickly find exact solutions~\cite{AndradeRW12,grosso2008simple}.

\subsubsection{Exact Algorithms.}
Much research has been devoted to reducing the base of the exponent for exact branch-and-bound algorithms. One main technique is to apply \emph{reductions}, which remove or modify subgraphs that can be solved simply, reducing the graph to a smaller instance. Reductions have consistently been used to reduce the running time of exact MIS algorithms~\cite{tarjan-1977}, with the current best polynomial-space algorithm having running time $O(1.2114^n)$~\cite{bourgeois-2012}. These algorithms apply reductions during recursion, only branching when the graph can no longer be reduced~\cite{fomin-2010}.

Relatively simple reduction techniques are known to be effective at reducing graph size in practice~\cite{abu-khzam-2007,butenko-2002}. Recently, Akiba and Iwata~\cite{akiba-tcs-2016} showed that more advanced reduction rules are also highly effective, finding an exact minimum vertex cover (and by extension, an exact maximum independent set) on a corpus of large social networks with up to 3.2 million vertices in less than a second.
However, their algorithm still requires $O(1.2210^n)$ time in the worst case, and its running time has exponential dependence on the kernel size. Since much larger graph instances have consistently large kernels, they remain intractable in practice~\cite{kamis-alenex-2016}.
Even though small benchmark graphs with up to thousands of vertices have been solved exactly with branch-and-bound algorithms~\cite{segundo-recoloring,segundo-bitboard-2011,tomita-recoloring}, many similarly-sized instances remain unsolved~\cite{butenko-2002}. Even a graph on 4,000 vertices was only recently solved exactly, and it required hundreds of machines in a MapReduce cluster~\cite{xiang-2013}. Heuristic algorithms are clearly still needed in practice, even for small graphs.

\subsubsection{Heuristic Approaches.}
\label{s:heuristic-approaches}
There are a wide range of heuristics and local search algorithms for the complementary maximum clique problem (see for example~\cite{battiti2001reactive,hansen2004variable,grosso2004combining,katayama2005effective,pullan2006dynamic,grosso2008simple}). 
These algorithms work by maintaining a single solution and attempt to improve it through node deletions, insertions, swaps, and \emph{plateau} search. Plateau search only accepts moves that do not change the objective function, which is typically achieved through \emph{node swaps}---replacing a node by one of its neighbors. Note that a node swap cannot directly increase the size of the independent set.
A very successful approach for the maximum clique problem has been presented by Grosso~\etal~\cite{grosso2008simple}. In addition to plateau search, it applies various diversification operations and restart rules. The iterated local search algorithm of Andrade~\etal\cite{AndradeRW12} is one of the most successful local search algorithms in practice. On small benchmark graphs requiring hours of computation to solve with exact algorithms, their algorithm often finds optimal solutions in milliseconds. However, for huge complex networks such as social networks and web graphs, it is consistently outperformed by other methods~\cite{lammSEA2015,kamis-alenex-2016}. We give further details of this algorithm in Section~\ref{s:ils}.

To solve these largest---and intractable---graphs, Lamm et al.~\cite{kamis-alenex-2016} proposed \redumis{}, an algorithm that uses reduction techniques combined with an evolutionary approach. It finds the exact MIS for many of the benchmarks used by Akiba et al.~\cite{akiba-tcs-2016}, and consistently finds larger independent sets than other heuristics.
Its major drawback is the significant preprocessing time it takes to apply reductions and initialize its evolutionary component, especially on larger instances. Thus, though \redumis{} finds high-quality independent sets faster than existing methods, it is still slow in practice on huge complex networks. However, for many of the applications mentioned above, a near-optimal independent set is not needed in practice. The main goal then is to quickly compute an independent set of sufficient quality. Hence, to find high-quality independent sets faster, we need a different approach.

\subsection{Our Results} 
We develop an advanced local search algorithm that quickly computes large independent sets by combining iterated local search with reduction rules that reduce the size of the search space without losing solution quality. 
By running local search on the kernel, we significantly boost its performance, especially on huge sparse networks.
In addition to exact kernelization techniques, we also apply inexact reductions that remove high-degree vertices from the graph.
In particular, we show that cutting a small percentage of high-degree vertices from the graph minimizes performance bottlenecks of local search, while maintaining high solution quality.
Experiments indicate that our algorithm finds large independent sets much faster than existing state-of-the-art algorithms, while still remaining competitive with the best solutions reported in literature. 

\section{Preliminaries}
\label{s:preliminaries}
Let  $G=(V=\{0,\ldots, n-1\},E)$ be an undirected graph with $n = |V|$ nodes and $m = |E|$ edges.
The set $N(v) = \setGilt{u}{\set{v,u}\in E}$ denotes the open neighborhood of $v$.
We further define the open neighborhood of a set of nodes $U \subseteq V$ to be $N(U) = \cup_{v\in U} N(v)\setminus U$.
We similarly define the closed neighborhood as $N[v] = N(v) \cup \{v\}$ and $N[U] = N(U) \cup U$.
A graph $H=(V_H, E_H)$ is said to be a \emph{subgraph} of $G=(V, E)$ if $V_H \subseteq V$ and $E_H \subseteq E$. We call $H$ an \emph{induced} subgraph when $E_H = \setGilt{\{u,v\} \in E}{u,v\in V_H}$.
For a set of nodes $U\subseteq V$, $G[U]$ denotes the subgraph induced by $U$.

An \emph{independent set} is a set $\mathcal{I} \subseteq V$, such that all nodes in $\mathcal{I}$ are pairwise nonadjacent. 
An independent set is \emph{maximal} if it is not a subset of any larger independent set.
The \emph{maximum independent set problem} is that of finding the maximum cardinality independent set among all possible independent sets. Such a set is called a \emph{maximum independent set} (MIS).

Finally, we note the maximum independent set problem is equivalent to the \emph{maximum clique} and \emph{minimum vertex cover} problems. We see this equivalence as follows: Given a graph $G=(V,E)$ and an independent set $\mathcal{I}\in V$, $V\setminus \mathcal{I}$ is a vertex cover and $I$ is a clique in the complement graph (the graph containing all edges missing in $G$). Thus, algorithms for any of these problems can also solve the maximum independent set problem.

\subsection{The ARW Algorithm}
\label{s:ils}
We now review the local search algorithm by Andrade~\etal\cite{AndradeRW12} (\arw) in more detail, since we use this algorithm in our work.
For the independent set problem, Andrade~\etal~\cite{AndradeRW12} extended the notion of swaps to $(j,k)$-swaps, which remove $j$ nodes from the current solution and insert $k$ nodes. The authors present a fast linear-time implementation that, given a maximal solution, can find a $(1,2)$-swap or prove that no $(1,2)$-swap exists. 
One \emph{iteration} of the \arw{} algorithm consists of a perturbation and a local search step.
The \arw{} \emph{local search} algorithm uses $(1,2)$-swaps to gradually improve a single current solution.  
The simple version of the local search iterates over all nodes of the graph and looks for a $(1,2)$-swap. By using a data structure that allows insertion and removal operations on nodes in time proportional to their degree, this procedure can find a valid $(1,2)$-swap in $\mathcal{O}(m)$ time, if it exists.

A \emph{perturbation step}, used for diversification, forces nodes into the solution and removes neighboring nodes as necessary.
In most cases a single node is forced into the solution; with a small probability the number of forced nodes $f$ is set to a higher value ($f$ is set to $i+1$ with probability $1/2^i$). Nodes to be forced into a solution are picked from a set of random candidates, with priority given to those that have been outside the solution for the longest time.
An even faster incremental version of the algorithm (which we use here) maintains a list of \emph{candidates}, which are nodes that may be involved in $(1,2)$-swaps. It ensures a node is not examined twice unless there is some change in its neighborhood. Furthermore, an external memory version of this algorithm by Liu~\etal~\cite{external-memory-ils} runs on graphs that do not fit into memory on a standard machine.
The \arw{} algorithm is efficient in practice, finding the exact maximum independent sets orders of magnitude faster than exact algorithms on many benchmark graphs.

\section{Techniques for Accelerating Local Search}
\label{s:noveltechniques}
\label{s:algorithms}
First, we note that while local search techniques such as \arw{} perform well on huge uniformly sparse mesh-like graphs, they perform poorly on complex networks, which are typically scale-free. We first discuss \emph{why} local search performs poorly on huge complex networks, then introduce the techniques we use to address these shortcomings.%--thus speeding up local search.

The first performance issue is related to vertex selection for perturbation. Many vertices are \emph{always} in some MIS. These include, for example, vertices with degree 1. However, \arw{} treats such vertices like any other. During a perturbation step, these vertices may be forced out of the current solution, causing extra searching that may not improve the solution.

The second issue is that high-degree vertices may slow \arw{} down significantly. Most internal operations of \arw{} (including (1,2)-swaps) require traversing the adjacency lists of multiple vertices, which takes time proportional to their degree. Although high-degree vertices are only scanned if they have at most one solution neighbor (or belong to the solution themselves), this happens often in complex networks.

A third issue is caused by the particular implementation.
When performing an (1,2)-swap involving the insertion of a vertex $v$, the original \arw{} implementation (as tested by Andrade~\etal\cite{AndradeRW12}) picks a pair of neighbors ${u,w}$ of $v$ at random among all valid ones. Although this technically violates that $O(m)$ worst-case bound (which requires the first such pair to be taken), the effect is minimal on the small-degree networks. On large complex networks, this can become a significant bottleneck.

To deal with the third issue, we simply modified the \arw{} code to limit the number of valid pairs considered to a small constant (100). Addressing the first two issues requires more involved techniques (\emph{kernelization} and \emph{high-degree vertex cutting}, respectively), as we discuss next.

\subsection{Exact Kernelization}
First, we investigate kernelization, a technique known to be effective in practice for finding an exact minimum vertex cover (and hence, a maximum independent set)~\cite{abu-khzam-2007,akiba-tcs-2016}. In kernelization, we repeatedly apply reductions to the input graph $G$ until it cannot be reduced further, producing a \emph{kernel} $\mathcal{K}$. Even simple reduction rules can significantly reduce the graph size, and in some cases $\mathcal{K}$ is empty---giving an exact solution without requiring any additional steps. We note that this is the case for many of the graphs in the experiments by Akiba and Iwata~\cite{akiba-tcs-2016}.
Furthermore, any solution of $\mathcal{K}$ can be extended to a solution of the input.

The size of the kernel depends entirely on the structure of the input graph. In many cases, the kernel can be too large, making it intractable to find an exact maximum independent set in practice (see Section~\ref{s:experiments}). 
In this case ``too large'' can mean a few thousand vertices. 
However, for many graphs, the kernel is still significantly smaller than the input graph, and even though it is intractable for exact algorithms, local search algorithms such as \arw{} have been shown to find the exact MIS quickly on small benchmark graphs. It therefore stands to reason that \arw{} would perform better on a small kernel.

\subsubsection{Reductions.}
We now briefly mention the reduction rules that we consider. Each of these exact reductions allow us to choose vertices that are in some MIS by following simple rules. If an MIS is found on the kernel graph $\mathcal{K}$, then each reduction may be undone, producing an MIS in the original graph. 

Akiba and Iwata~\cite{akiba-tcs-2016} use a full suite of reduction rules, which they show can efficiently solve the minimum vertex cover problem exactly for a wide variety of instances. We consider all of their reductions here. These include simple rules typically used in practice such as \emph{pendant vertex removal} and \emph{vertex folding}~\cite{chen-1999}, and more advanced (and time-consuming) rules such as a linear programming relaxation with a half-integral solution~\cite{iwata-2014,nemhauser-1975}, \emph{unconfined}~\cite{Xiao201392}, \emph{twin}~\cite{Xiao201392}, \emph{alternative}~\cite{Xiao201392}, and \emph{packing}~\cite{akiba-tcs-2016} reductions. Since details on these reductions are not necessary for understanding our results, 
\begin{wrapfigure}[10]{r}{2.75cm}
\centering
\vspace{-.75cm}
\includegraphics[width=2.25cm]{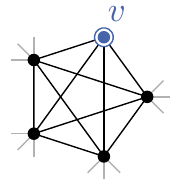}
\caption{An isolated vertex $v$, in a single clique of five vertices.}
\end{wrapfigure}
we defer them to Appendix~\ref{sec:reductions}.
(Refer to Akiba and Iwata~\cite{akiba-tcs-2016} for a more thorough discussion, including implementation details.)

The most relevant reduction for our purposes is the \emph{isolated vertex removal}. If a vertex $v$ forms a single clique $C$ with all its neighbors, then $v$ is called \emph{isolated} and is always contained in some maximum independent set.
To see this, at most one vertex from $C$ may be in an MIS. If a neighbor of $v$ is in an MIS, then so is $v$. Otherwise, $v$ is in the MIS. This reduction was not included in the exact algorithm by Akiba and Iwata~\cite{akiba-tcs-2016}; however, Butenko~\etal~\cite{butenko-2002} show that it is highly effective on graphs derived from error-correcting codes.

\subsection{Inexact Reductions: Cutting High-Degree Vertices}
To further boost local search, we investigate removing (cutting) high-degree vertices outright. This is a natural strategy:
intuitively, vertices with very high degree are unlikely to be in a large independent set (consider a maximum independent set of graphs with few high-degree vertices, such as a star graph, or scale-free networks). In particular, many reduction rules show that low-degree vertices are in some MIS, and applying them results in a small kernel~\cite{kamis-alenex-2016}. Thus, high-degree vertices are left behind. This is especially true for huge complex networks considered here, which generally have few high-degree vertices.

Besides intuition, there is much additional evidence to support this strategy. In particular, the natural greedy algorithm that repeatedly selects low-degree vertices to construct an independent set is typically within 1\%--10\% of the maximum independent set size for sparse graphs~\cite{AndradeRW12}. Moreover, several successful algorithms make choices that favor low-degree vertices. \redumis{}~\cite{kamis-alenex-2016} forces low-degree vertices into an independent set in a multi-level algorithm, giving high-quality independent sets as a result. Exact branch-and-bound algorithms order vertices so that vertices of high-degree are considered first during search. This reduces the search space size initially, at the cost of finding poor initial independent sets. In particular, optimal and near-optimal independent sets are typically found after high-degree vertices have been evaluated and excluded from search;  however, it is then much slower to find the remaining solutions, since only low-degree vertices remain in the search. This slowness can be observed in the experiments of Batsyn et al.~\cite{batsyn-mcs-ils-2014}, where better initial solutions from local search significantly speed up exact search.

%\subsubsection{Absolute vs. Relative Degree Cutting.}

%A $k$-core decomposition iteratively removes the minimum degree vertex to compute the graph's $k$-core. Since we want to cut the highest degree vertices, we therefore perform the $k$-core decomposition on the graph's complement. Thus, we achieve the complement of a $k$-core in the primal graph which only contains vertices of degree less than or equal to $|V|-k$. For sparse primal graphs $k$ is close to $|V|$. 
%Note that we do not have to explicitly compute the complement to select the vertices to remove from graph.
%To perform the cutting, we use a bucket priority queue that maintains all vertices in buckets according to their degrees.

We considered two strategies for removing high-degree vertices from the graph.
\csch{currently not clear how we break ties}
When we cut by \emph{absolute degree}, we remove the vertices with degree higher than a threshold. In \emph{relative degree} cutting, we iteratively remove high-degree vertices and their incident edges from the graph. This is the mirror image of the greedy algorithm that repeatedly selects smallest-degree vertices in the graph to be in an independent set until the graph is empty. We stop removing until a fixed fraction of all vertices is eliminated. Unlike absolute cutting, this better ensures that clusters of high-degree vertices are removed, leaving high-degree vertices that are isolated from one another, which are more likely to be in large independent sets.

\subsection{Putting Things Together}
We use reductions and cutting in two ways. First, we define an algorithm that applies the standard technique of producing the kernel in advance, and then run \arw{} on the kernel. Second, we investigate applying reductions online as \arw{} runs.

\vspace{-.3cm}
\subsubsection{Preprocessing.}
Our first algorithm (\kermis) uses exact reductions in combination with relative degree cutting.
It uses the full set of reductions from Akiba and Iwata~\cite{akiba-tcs-2016}, as described in Section~\ref{s:algorithms}. Note that we do not include isolated vertex removal, as it was not included in their reductions.
After computing a kernel, we then cut 1\% of the highest-degree vertices using relative cutting, breaking ties randomly.
We then run \arw{} on the resulting graph.
% A more advanced tie-breaking mechanism, that factors in the sum of neighbors' degrees is also tested.
% This variant always chooses the vertex with the highest sum of its neighbors' degrees, between all vertices with the same degree.
% However, our experiments with this technique did not yield better results so that we omit from further discussions.

\vspace{-.3cm}
\subsubsection{Online.}
Our second approach (\online) applies a set of simple reductions on the fly.
For this algorithm, we use only the isolated vertex removal reduction (for degree zero, one and two), since it does not require the graph to be modified---we can just mark isolated vertices and their neighbors as removed during local search.
In more detail, we first perform a quick \emph{single pass} when computing the initial solution for \arw{}. We force isolated vertices into the initial solution, and mark them and their neighbors as removed. Note that this does not result in a kernel, as this pass may create more isolated vertices. We further mark the top 1\% of high-degree vertices as removed during this pass.
As the local search continues, whenever we check if a vertex can be inserted into the solution, we check if it is isolated and update the solution and graph similarly to the single pass. Thus, \online{} kernelizes the graph in an online fashion as local search proceeds.

\section{Experimental Evaluation}

\label{s:experiments}
\subsection{Methodology} 
We implemented our algorithms (\online, \kermis) including the kernelization techniques using C++ and compiled all code using gcc 4.63 with full optimizations turned on (\texttt{-O3} flag). 
We further compiled the original implementations of \arw{} and \redumis{} using the same settings. For \redumis, we use the same parameters as Lamm~\etal~\cite{kamis-alenex-2016} (convergence parameter $\mu=1,000,000$, reduction parameter $\lambda=0.1\cdot|\mathcal{I}|$, and cutting percentage $\eta=0.1 \cdot |\mathcal{K}|$).
For all instances, we perform three independent runs of each algorithm. For small instances, we run each algorithm sequentially with a 5-hour wall-clock time limit to compute its best solution. For huge graphs, with tens of millions of vertices and at least one billion edges, we enforce a time limit of ten hours.

Each run was performed on a machine that is equipped with four Octa-Core Intel Xeon E5-4640 processors running at 2.4\,GHz.
It has 512 GB local memory, $4\times 20$ MB L3-Cache and $4\times 8\times256$ KB L2-Cache.

We consider social networks, autonomous systems graphs, and Web graphs taken from the 10th DIMACS Implementation Challenge~\cite{benchmarksfornetworksanalysis}, and two additional large Web graphs, \Id{webbase-2001}~\cite{webgraphWS} and \Id{wikilinks}~\cite{kunegis-2013}. 
We also include road networks from Andrade~\etal~\cite{AndradeRW12} and meshes from Sander~\etal~\cite{sander-mesh-2008}.
The graphs \Id{europe} and \Id{USA-road} are large road networks of Europe~\cite{DSSW09} and the USA~\cite{demetrescu2009shortest}. 
The instances \Id{as-Skitter-big}, \Id{web-Stanford} and \Id{libimseti} are the hardest instances from Akiba and Iwata~\cite{akiba-tcs-2016}. 
We further perform experiments on huge instances with billions of edges taken from the Laboratory of Web Algorithmics~\cite{webgraphWS}:  \Id{it-2004}, \Id{sk-2005}, and \Id{uk-2007}.

\subsection{Accelerated Solutions}

\begin{table}[!tb]
\centering
\caption{ For each graph instance, we give the number of vertices $n$ and the number of edges $m$. We further give the maximum speedup for \online{} over other heuristic search algorithms. For each solution size $i$, we compute the speedup $s_{\AlgName{Alg}}^i = t_{\AlgName{Alg}}^i/t_{\online}^i$ of \online{} over algorithm \AlgName{Alg} for that solution size. We then report the maximum speedup $s_{\AlgName{Alg}}^{\max}=\max_i{s_{\AlgName{Alg}}^i}$ for the instance. When an algorithm never matches the final solution quality of \online, we give the highest non-infinite speedup and give an *. A `$\infty$' indicates that all speedups are infinite.}
\label{table:speedup}
\scriptsize
%\begin{tabulary}{\textwidth}{l r r r r r}
%\hline
%\multicolumn{3}{c}{Graph} & \multicolumn{3}{c}{Max. speedup of \online{} vs\ldots} \\
%\hline
%Name & $n$ & $m$ & \arw & \kermis & \redumis \\ 
%\hline
\detailedheadertabspeedup
\multicolumn{6}{l}{\bf Huge instances:}\\[0.2em]
\Id{\detokenize{it-2004}} & \numprint{41291594} & \numprint{1027474947} & \numprint{4.51}\phantom{*}& \numprint{221.26}\phantom{*}& \numprint{266.30}\phantom{*}\\ 
 \Id{\detokenize{sk-2005}} & \numprint{50636154} & \numprint{1810063330} & \numprint{356.87}* & \numprint{201.68}\phantom{*}& \numprint{302.64}\phantom{*}\\
  \Id{\detokenize{uk-2007}} & \numprint{105896555} & \numprint{1154392916} & \numprint{11.63}* & \numprint{108.13}\phantom{*}& \numprint{122.50}\phantom{*}\\
\multicolumn{6}{l}{\bf Social networks and Web graphs:}\\[0.2em]
\Id{\detokenize{amazon-2008}} & \numprint{735323} & \numprint{3523472} & \numprint{43.39}* & \numprint{13.75}\phantom{*}& \numprint{50.75}\phantom{*}\\ 
 \Id{\detokenize{as-Skitter-big}} & \numprint{1696415} & \numprint{11095298} & \numprint{355.06}* & \numprint{2.68}\phantom{*}& \numprint{7.62}\phantom{*}\\ 
 \Id{\detokenize{dewiki-2013}} & \numprint{1532354} & \numprint{33093029} & \numprint{36.22}* & \numprint{632.94}\phantom{*}& \numprint{1726.28}\phantom{*}\\ 
 \Id{\detokenize{enwiki-2013}} & \numprint{4206785} & \numprint{91939728} & \numprint{51.01}* & \numprint{146.58}\phantom{*}& \numprint{244.64}\phantom{*}\\ 
 \Id{\detokenize{eu-2005}} & \numprint{862664} & \numprint{22217686} & \numprint{5.52}\phantom{*}& \numprint{62.37}\phantom{*}& \numprint{217.39}\phantom{*}\\ 
 \Id{\detokenize{hollywood-2011}} & \numprint{2180759} & \numprint{114492816} & \numprint{4.35}\phantom{*}& \numprint{5.51}\phantom{*}& \numprint{11.24}\phantom{*}\\ 
 \Id{\detokenize{libimseti}} & \numprint{220970} & \numprint{17233144} & \numprint{15.16}* & \numprint{218.30}\phantom{*}& \numprint{1118.65}\phantom{*}\\ 
 \Id{\detokenize{ljournal-2008}} & \numprint{5363260} & \numprint{49514271} & \numprint{2.51}\phantom{*}& \numprint{3.00}\phantom{*}& \numprint{5.33}\phantom{*}\\ 
 \Id{\detokenize{orkut}} & \numprint{3072441} & \numprint{117185082} & \numprint{1.82}* & \numprint{478.94}* & \numprint{8751.62}* \\ 
 \Id{\detokenize{web-Stanford}} & \numprint{281903} & \numprint{1992636} & \numprint{50.70}* & \numprint{29.53}\phantom{*}& \numprint{59.31}\phantom{*}\\ 
 \Id{\detokenize{webbase-2001}} & \numprint{118142155} & \numprint{854809761} & \numprint{3.48}\phantom{*}& \numprint{33.54}\phantom{*}& \numprint{36.18}\phantom{*}\\ 
 \Id{\detokenize{wikilinks}} & \numprint{25890800} & \numprint{543159884} & \numprint{3.88}\phantom{*}& \numprint{11.54}\phantom{*}& \numprint{11.89}\phantom{*}\\ 
 \Id{\detokenize{youtube}} & \numprint{1134890} & \numprint{543159884} & \numprint{6.83}\phantom{*}& \numprint{1.83}\phantom{*}& \numprint{7.29}\phantom{*}\\ 
\multicolumn{6}{l}{\bf Road networks:}\\[0.2em]
 \Id{\detokenize{europe}} & \numprint{18029721} & \numprint{22217686} & \numprint{5.57}\phantom{*}& \numprint{12.79}\phantom{*}& \numprint{14.20}\phantom{*}\\ 
\Id{\detokenize{USA-road}} & \numprint{23947347} & \numprint{28854312} & \numprint{7.17}\phantom{*}& \numprint{24.41}\phantom{*}& \numprint{27.84}\phantom{*}\\ 
\multicolumn{6}{l}{\bf Meshes:}\\[0.2em]
\Id{\detokenize{buddha}} & \numprint{1087716} & \numprint{1631574} & \numprint{1.16}\phantom{*}& \numprint{154.04}* & \numprint{976.10}* \\ 
 \Id{\detokenize{bunny}} & \numprint{68790} & \numprint{103017} & \numprint{3.26}\phantom{*}& \numprint{16616.83}* & \numprint{526.14}\phantom{*}\\ 
 \Id{\detokenize{dragon}} & \numprint{150000} & \numprint{225000} & \numprint{2.22}* & \numprint{567.39}* & \numprint{692.60}* \\ 
 \Id{\detokenize{feline}} & \numprint{41262} & \numprint{61893} & \numprint{2.00}* & \numprint{13377.42}* & \numprint{315.48}\phantom{*}\\ 
 \Id{\detokenize{gameguy}} & \numprint{42623} & \numprint{63850} & \numprint{3.23}\phantom{*}& \numprint{98.82}* & \numprint{102.03}\phantom{*}\\ 
 \Id{\detokenize{venus}} & \numprint{5672} & \numprint{8508} & \numprint{1.17}\phantom{*}& $\infty$\phantom{*} & \numprint{157.78}* \\ 
 \bottomrule
\end{tabulary}
\vspace*{-.5cm}
\end{table}

We now illustrate the speed improvement over existing heuristic algorithms. First, we measure the speedup of \online{} over other high-quality heuristic search algorithms. In particular, in Table~\ref{table:speedup}, we report the maximum speedup that \online{} compared with the state-of-the-art competitors. We compute the maximum speedup for an instance as follows. For each solution size $i$, we compute the speedup $s_{\AlgName{Alg}}^i = t_{\AlgName{Alg}}^i/t_{\online}^i$ of \online{} over algorithm \AlgName{Alg} for that solution size. We then report the maximum speedup $s_{\AlgName{Alg}}^{\max}=\max_i{s_{\AlgName{Alg}}^i}$ for the instance. 

As can be seen in Table~\ref{table:speedup}, \online{} always has a maximum speedup greater than 1 over every other algorithm. We first note that \online{} is significantly faster than \redumis{} and \kermis{}. In particular, in 14 instances, \online{} achieves a maximum speedup of over 100 over \redumis{}. \kermis{} performs only slightly better than \redumis{} in this regard, with \online{} achieving similar speedups on 12 instances. Though, on meshes, \kermis{} fairs especially poorly. On these instances, \online{} always finds a better solution than \kermis{} (instances marked with an *), and on the \Id{bunny} and \Id{feline} instances, \online{} achieves a maximum speedup of over 10,000 against \kermis{}. Furthermore, on the \Id{venus} mesh graph, \kermis{} never matches the quality of a single solution from \online{}, giving infinite speedup.  
\arw{} is the closest competitor, where \online{} only has 2 maximum speedups greater than 100. However, on a further 6 instances, \online{} achieves a maximum speedup over 10, and on 11 instances \arw{} fails to match the final solution quality of \online{}, giving an effective infinite maximum speedup.

\begin{figure}[!t]
\centering
\includegraphics[width=0.49\textwidth]{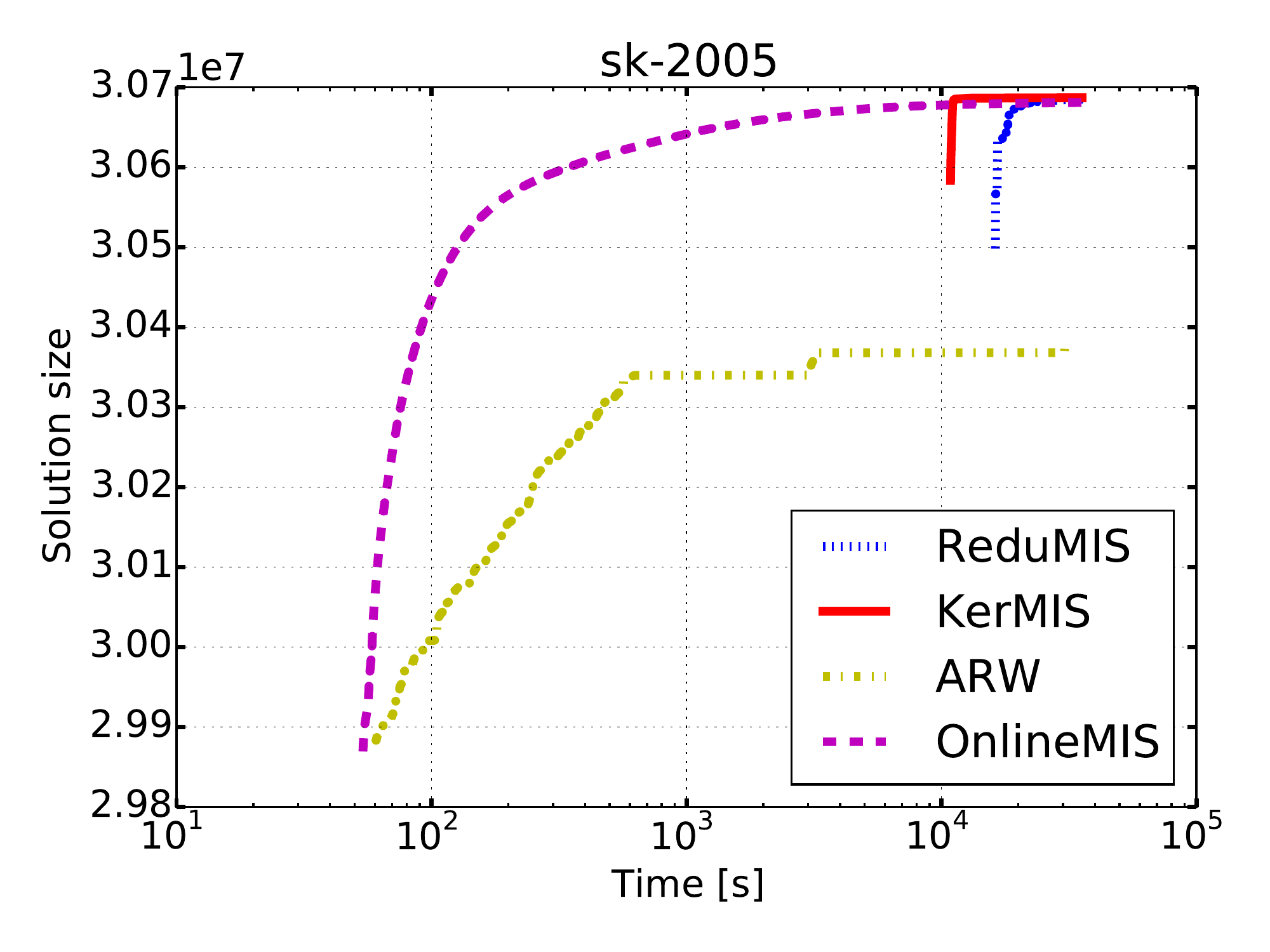}
\includegraphics[width=0.49\textwidth]{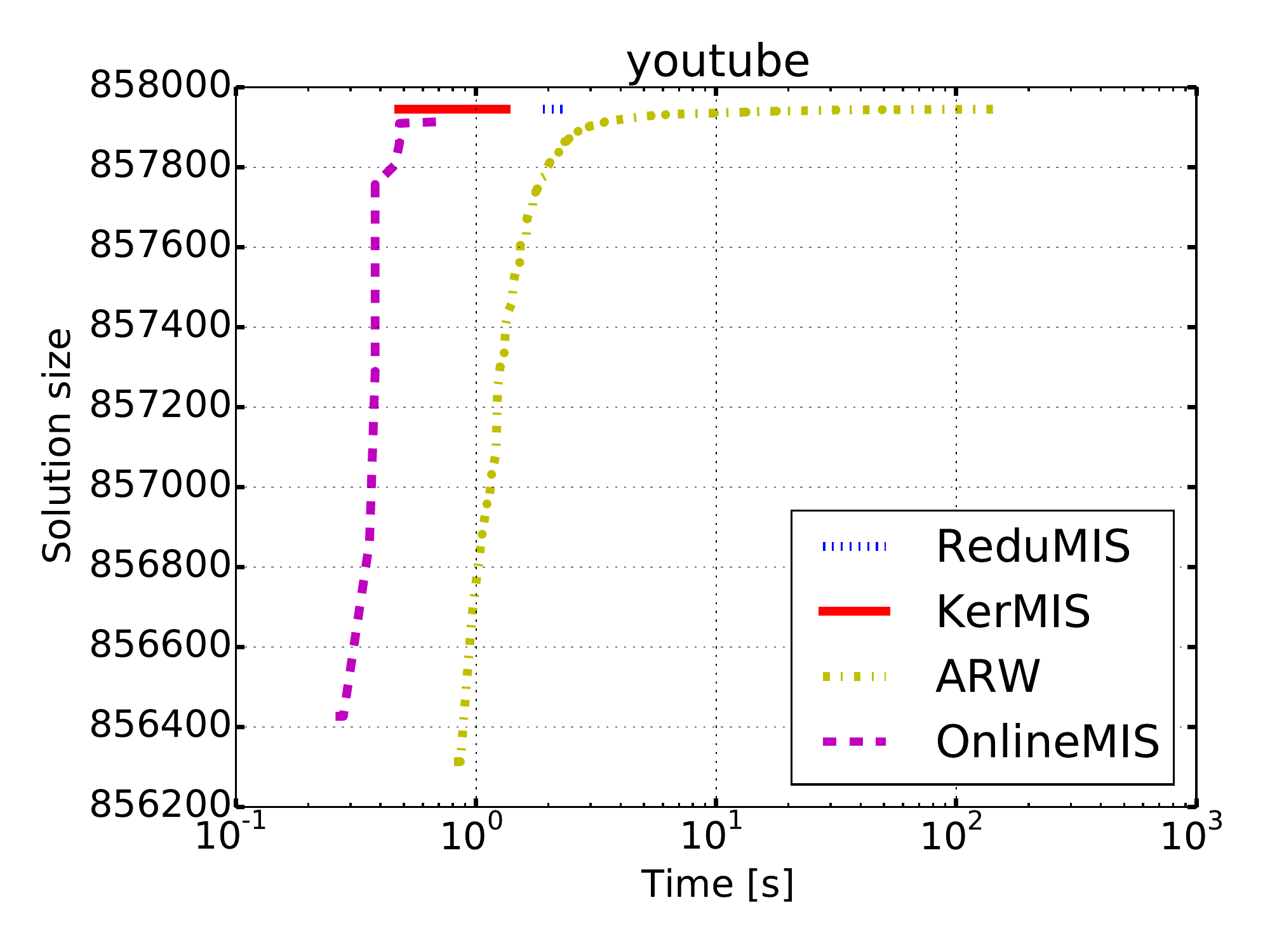} \\
\includegraphics[width=0.49\textwidth]{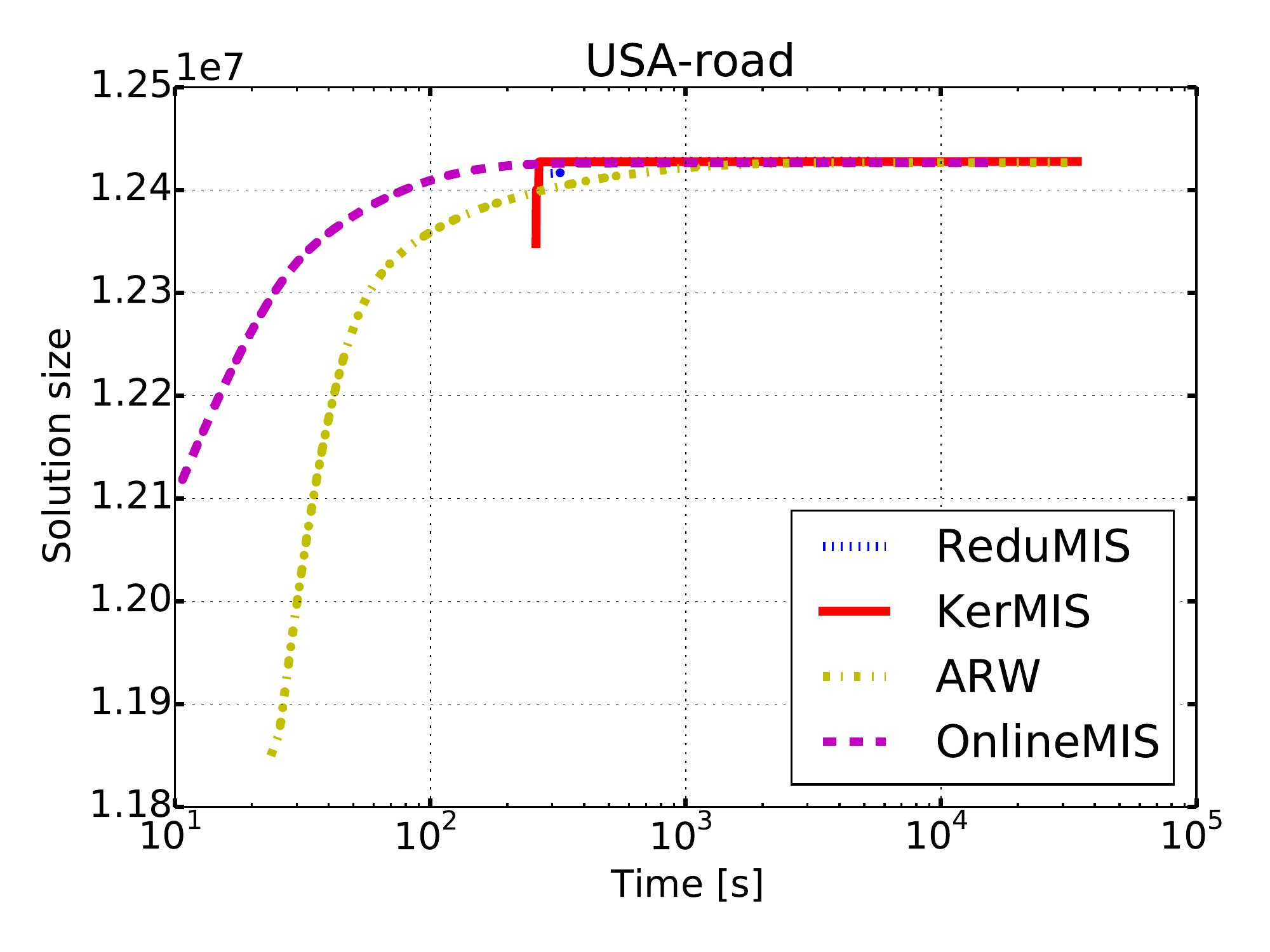}
\includegraphics[width=0.49\textwidth]{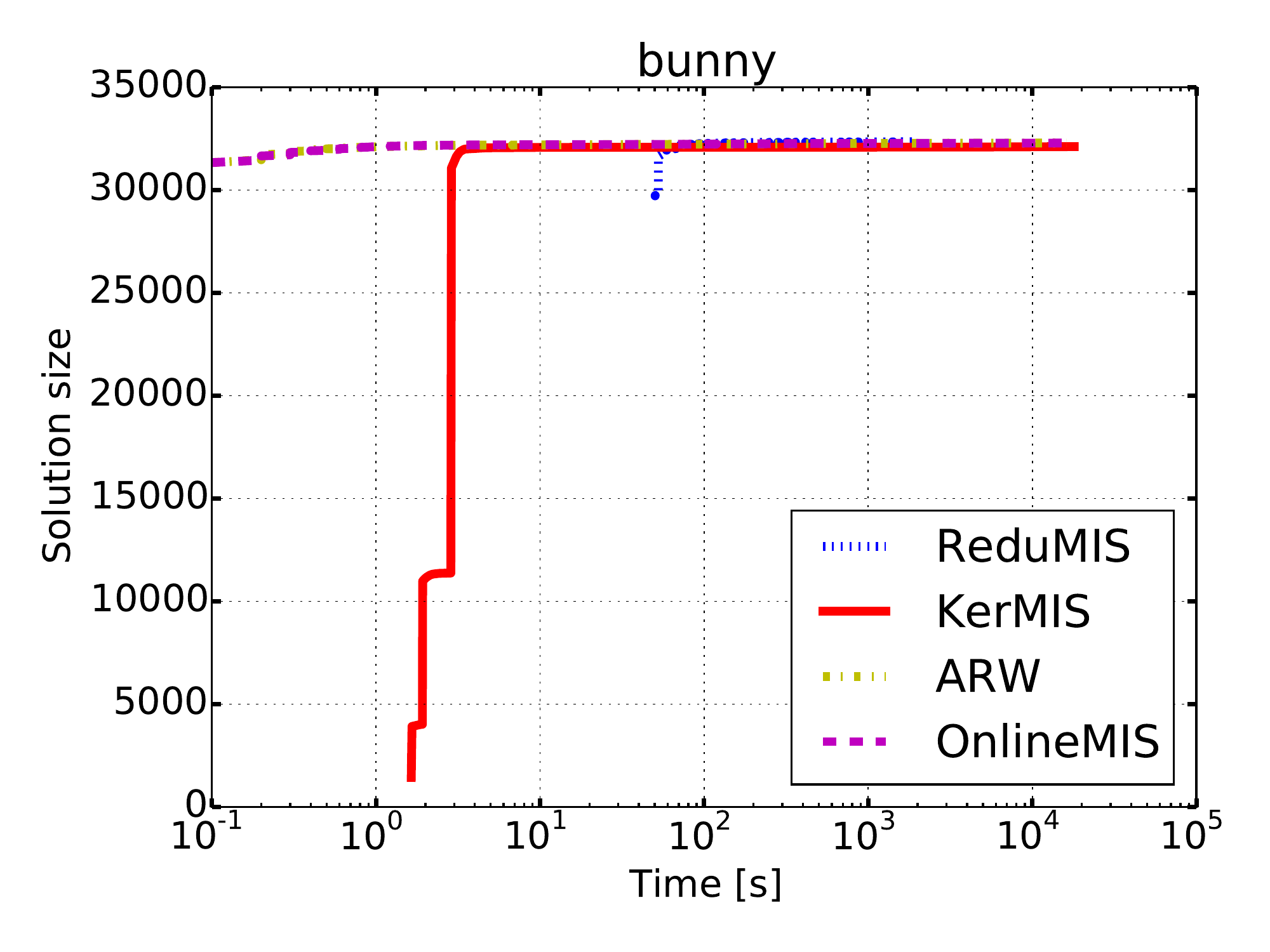}
\caption{Convergence plots for \Id{\detokenize{sk-2005}} (top left), \Id{\detokenize{youtube}} (top right), \Id{\detokenize{USA-road}} (bottom left), and \Id{\detokenize{bunny}} (bottom right). \csch{unify the y-axis in the plots? can we insert a small window showing the algorithms at the end to see the difference? why is there two times USA?}\strash{We can insert a window into the latter part of the plots, but is it really needed?} \csch{just meant the formatting of the numbers ... 1e7 is somehow strange}}
\label{fig:convergenceplots}
\vspace*{-.5cm}
\end{figure}

We now give several representative \emph{convergence plots} in Fig.~\ref{fig:convergenceplots}, which illustrate the early solution quality of \online{} compared to \arw, the closest competitor. We construct these plots as follows. Whenever an algorithm finds a new large independent set $I$ at time $t$, it reports a tuple ($t$, $|I|$); the convergence plots show average values over all three runs. In the non-mesh instances, \online{} takes a early lead over \arw{}, though solution quality converges over time. Lastly, we give the convergence plot for the \Id{bunny} mesh graph. Reductions and high-degree cutting aren't effective on meshes, thus \arw{} and \online{} have similar initial solution sizes.

\subsection{Time to High-Quality Solutions}
\begin{table}[tb]
\centering
\caption{For each algorithm, we give the average time $t_{avg}$ to reach 99.5\% of the best solution found by any algorithm. The fastest such time for each instance is marked in bold. We also give the size of the largest solution found by any algorithm and list the algorithms (abbreviated by first letter) that found this largest solution in the time limit. A `-' indicates that the algorithm did not find a solution of sufficient quality.}
\label{table:percent}
\scriptsize
\hspace*{-.25cm}
\detailedheadertabpercent
\multicolumn{7}{l}{\bf Huge instances:}\\[0.2em]
\Id{\detokenize{it-2004}}                                        & \textbf{\numprint{86.01}}  & \numprint{327.35}        & \numprint{7892.04}  & \numprint{9448.18}  & \numprint{25620285} & \phantom{\AlgName{O, }}\phantom{\AlgName{A, }}\phantom{\AlgName{K}, }\AlgName{R} \\
\Id{\detokenize{sk-2005}}                                        & \textbf{\numprint{152.12}} & -                        & \numprint{10854.46} & \numprint{16316.59} & \numprint{30686766} & \phantom{\AlgName{O, }}\phantom{\AlgName{A, }}\AlgName{K}\phantom{\AlgName{, R}} \\
 \Id{\detokenize{uk-2007}} & \textbf{\numprint{403.36}} & \numprint{3789.74} & \numprint{23022.26} & \numprint{26081.36} & \numprint{67282659} & \phantom{\AlgName{O, }}\phantom{\AlgName{A, }}\AlgName{K}\phantom{\AlgName{, R}} \\
\multicolumn{7}{l}{\bf Social networks and Web graphs:}\\[0.2em]
%\hline
\Id{\detokenize{amazon-2008}}                                    & \textbf{\numprint{0.76}}   & \numprint{1.26}          & \numprint{5.81}     & \numprint{15.23}    & \numprint{309794}   & \phantom{\AlgName{O, }}\phantom{\AlgName{A, }}\AlgName{K}, \AlgName{R} \\
\Id{\detokenize{as-Skitter-big}}                                 & \textbf{\numprint{1.26}}   & \numprint{2.70}           & \numprint{2.82}     & \numprint{8.00}      & \numprint{1170580}  & \phantom{\AlgName{O, }}\phantom{\AlgName{A, }}\AlgName{K}, \AlgName{R} \\
\Id{\detokenize{dewiki-2013}}                                    & \textbf{\numprint{4.10}}    & \numprint{7.88}          & \numprint{898.77}   & \numprint{2589.32}  & \numprint{697923}   & \phantom{\AlgName{O, }}\phantom{\AlgName{A, }}\AlgName{K}\phantom{, \AlgName{R}} \\
\Id{\detokenize{enwiki-2013}}                                    & \textbf{\numprint{10.49}}  & \numprint{19.26}         & \numprint{856.01}   & \numprint{1428.71}  & \numprint{2178457}  & \phantom{\AlgName{O, }}\phantom{\AlgName{A, }}\AlgName{K}\phantom{, \AlgName{R}} \\
\Id{\detokenize{eu-2005}}                                        & \textbf{\numprint{1.32}}   & \numprint{3.11}          & \numprint{29.01}    & \numprint{95.65}    & \numprint{452353}   & \phantom{\AlgName{O, }}\phantom{\AlgName{A, }}\phantom{\AlgName{K}, }\AlgName{R} \\
\Id{\detokenize{hollywood-2011}}                                 & \textbf{\numprint{1.28}}   & \numprint{1.46}          & \numprint{7.06}     & \numprint{14.38}    & \numprint{523402}   & \AlgName{O}, \AlgName{A}, \AlgName{K}, \AlgName{R} \\
\Id{\detokenize{libimseti}}                                      & \textbf{\numprint{0.44}}   & \numprint{0.45}          & \numprint{50.21}    & \numprint{257.29}   & \numprint{127293}   & \phantom{\AlgName{O, }}\phantom{\AlgName{A, }}\phantom{\AlgName{K}, }\AlgName{R} \\
\Id{\detokenize{ljournal-2008}}                                  & \textbf{\numprint{3.79}}   & \numprint{8.30}           & \numprint{10.20}     & \numprint{18.14}    & \numprint{2970937}  & \phantom{\AlgName{O, }}\phantom{\AlgName{A, }}\AlgName{K}, \AlgName{R} \\
\Id{\detokenize{orkut}}                                          & \textbf{\numprint{42.19}}  & \numprint{49.18}         & \numprint{2024.36}  & -                   & \numprint{839086}   & \phantom{\AlgName{O, }}\phantom{\AlgName{A, }}\AlgName{K}\phantom{, \AlgName{R}} \\
\Id{\detokenize{web-Stanford}}                                   & \textbf{\numprint{1.58}}   & \numprint{8.19}          & \numprint{3.57}     & \numprint{7.12}     & \numprint{163390}   & \phantom{\AlgName{O, }}\phantom{\AlgName{A, }}\phantom{\AlgName{K}, }\AlgName{R} \\
\Id{\detokenize{webbase-2001}}                                   & \textbf{\numprint{144.51}} & \numprint{343.86}        & \numprint{2920.14}  & \numprint{3150.05}  & \numprint{80009826} & \phantom{\AlgName{O, }}\phantom{\AlgName{A, }}\phantom{\AlgName{K}, }\AlgName{R} \\
\Id{\detokenize{wikilinks}}                                    & \textbf{\numprint{34.40}}   & \numprint{85.54}         & \numprint{348.63}   & \numprint{358.98}   & \numprint{19418724} & \phantom{\AlgName{O, }}\phantom{\AlgName{A, }}\phantom{\AlgName{K}, }\AlgName{R} \\
\Id{\detokenize{youtube}}                                        & \textbf{\numprint{0.26}}   & \numprint{0.81}          & \numprint{0.48}     & \numprint{1.90}      & \numprint{857945}   & \phantom{\AlgName{O, }}\AlgName{A}, \AlgName{K}, \AlgName{R} \\
\multicolumn{7}{l}{\bf Road networks:}\\[0.2em]
%\hline
\Id{\detokenize{europe}}                                            & \textbf{\numprint{28.22}}  & \numprint{75.67}         & \numprint{91.21}    & \numprint{101.21}   & \numprint{9267811}  & \phantom{\AlgName{O, }}\phantom{\AlgName{A, }}\phantom{\AlgName{K}, }\AlgName{R} \\
\Id{\detokenize{USA-road}}                                       & \textbf{\numprint{44.21}}  & \numprint{112.67}        & \numprint{259.33}   & \numprint{295.70}    & \numprint{12428105} & \phantom{\AlgName{O, }}\phantom{\AlgName{A, }}\phantom{\AlgName{K}, }\AlgName{R} \\
%\hline
\multicolumn{7}{l}{\bf Meshes:}\\[0.2em]
\Id{\detokenize{buddha}}                                         & \textbf{\numprint{26.23}}  & \numprint{26.72}         & \numprint{119.05}   & \numprint{1699.19}  & \numprint{480853}   & \phantom{\AlgName{O, }}\AlgName{A}\phantom{\AlgName{K}, }\phantom{\AlgName{R}} \\
\Id{\detokenize{bunny}}                                          & \textbf{\numprint{3.21}}   & \numprint{9.22}          & -                   & \numprint{70.40}     & \numprint{32349}    & \phantom{\AlgName{O, }}\phantom{\AlgName{A}, }\phantom{\AlgName{K}, }\AlgName{R} \\
\Id{\detokenize{dragon}}                                         & \textbf{\numprint{3.32}}   & \numprint{4.90}           & \numprint{5.18}     & \numprint{97.88}    & \numprint{66502}    & \phantom{\AlgName{O, }}\AlgName{A}\phantom{\AlgName{K}, } \\
\Id{\detokenize{feline}}                                         & \textbf{\numprint{1.24}}   & \numprint{1.27}          & -                   & \numprint{39.18}    & \numprint{18853}    & \phantom{\AlgName{O, }}\phantom{\AlgName{A}, }\phantom{\AlgName{K}, }\AlgName{R} \\
\Id{\detokenize{gameguy}}                                        & \numprint{15.13}           & \textbf{\numprint{10.60}} & \numprint{60.77}    & \numprint{12.22}    & \numprint{20727}    & \phantom{\AlgName{O, }}\phantom{\AlgName{A}, }\phantom{\AlgName{K}, }\AlgName{R} \\
\Id{\detokenize{venus}}                                          & \textbf{\numprint{0.32}}   & \numprint{0.36}          & -                   & \numprint{6.52}     & \numprint{2684}     & \AlgName{O}, \AlgName{A}, \phantom{\AlgName{K}, }\AlgName{R} \\
%\hline
 \bottomrule
\end{tabulary}
 \vspace*{-.5cm}
\end{table}

We now look at the time it takes an algorithm to find a high-quality solution. We first determine the largest independent set found by any of the four algorithms, which represent the best-known solutions~\cite{kamis-alenex-2016}, and compute how long it takes each algorithm to find an independent set within $99.5$\% of this size.
The results are shown in Table~\ref{table:percent}. With a single exception, \online{} is the fastest algorithm to be within $99.5$\% of the target solution. In fact, \online{} finds such a solution at least twice as fast as \arw{} in $14$ instances; it is almost 10 times faster on the largest instance, \Id{uk-2007}. Further, \online{} is orders of magnitude faster than \redumis{} (by a factor of at least 100 in 7 cases). We also see that \kermis{} is faster than \redumis{} in 19 cases, but much slower than \online{} for all instances. It does eventually find the largest independent set (among all algorithms) for 10 instances. This shows that the full set of reductions is not always necessary, especially when the goal is to get a high-quality solution quickly. It also justifies our choice of cutting: the solution quality of \kermis{} rivals (and sometimes even improves) that of \redumis.
\ifFull

\subsection{Overall Solution Quality}
Next, we show that \online{} has high solution quality when given a time limit for searching ($5$ hours for normal graphs, $10$ hours for huge graphs).
Although long-run quality is not the goal of the \online{} algorithm, in $11$ instances \online{} finds a larger independent set than \arw, and in $4$ instances \online{} finds the largest solution in the time limit.
As seen in Table~\ref{table:avg}, \online{} also finds a solution within $0.1$\% of the best solution found by any algorithm for all graphs.
However, in general \online{} finds lower-quality solutions than \redumis, which we believe is from high-degree
cutting removing vertices in large independent sets.
Nonetheless, as this shows, even when cutting out $1$\% of the vertices, the solution quality remains high.
\begin{table}[!tb]
\centering
\caption{For each algorithm, we include average solution size and average time $t_{avg}$ to reach it within a time limit (5 hours for normal graphs, 10 hours for huge graphs). Solutions in italics indicate the larger solution between \arw{} and \online{} local search, bold marks the largest overall solution. A `-' in our indicates that the algorithm did not find a solution in the time limit.}
\label{table:avg}
\scriptsize
\hspace*{-.25cm}
\detailedheadertabavg
%\begin{tabulary}{\textwidth}{l|r r|r r|r r|r r}
%\hline
%\multicolumn{1}{c}{Graph} & \multicolumn{2}{|c|}{\online} & \multicolumn{2}{|c}{\arw} & \multicolumn{2}{|c|}{\kermis} & \multicolumn{2}{|c|}{\redumis} \\
%\hline
%Name & Avg. & $t_{avg}$ & Avg. & $t_{avg}$ & Avg. & $t_{avg}$ & Avg. & $t_{avg}$ \\[0.8em]
%\hline 
\multicolumn{9}{l}{\bf Huge instances:}\\[0.2em]
%\hline
\Id{\detokenize{it-2004}}                                        & \numprint{25610697}                 & \numprint{35324} & \numprint{25612993}        & \numprint{33407} & \numprint{25619988}          & \numprint{35751} & \textbf{\numprint{25620246}} & \numprint{35645} \\
\Id{\detokenize{sk-2005}}                                        & \textit{\numprint{30680869}}        & \numprint{34480} & \numprint{30373880}        & \numprint{11387} & \textbf{\numprint{30686684}} & \numprint{34923} & \numprint{30684867}          & \numprint{35837} \\
 \Id{\detokenize{uk-2007}} & \textit{\numprint{67265560}} & \numprint{35982} & \numprint{67101065} & \numprint{8702} & \textbf{\numprint{67282347}} & \numprint{35663} & \numprint{67278359} & \numprint{35782} \\
%\hline
\multicolumn{9}{l}{\bf Social networks and Web graphs:}\\[0.2em]
%\hline
\Id{\detokenize{amazon-2008}}                                    & \textit{\numprint{309792}}          & \numprint{6154}  & \numprint{309791}          & \numprint{12195} & \numprint{309793}            & \numprint{818}   & \textbf{\numprint{309794}}   & \numprint{153} \\
\Id{\detokenize{as-Skitter-big}}                                 & \textit{\numprint{1170560}}         & \numprint{7163}  & \numprint{1170548}         & \numprint{14017} & \textbf{\numprint{1170580}}  & \numprint{4}     & \textbf{\numprint{1170580}}  & \numprint{9} \\
\Id{\detokenize{dewiki-2013}}                                    & \textit{\numprint{697789}}          & \numprint{17481} & \numprint{697669}          & \numprint{16030} & \textbf{\numprint{697921}}   & \numprint{14070} & \numprint{697798}            & \numprint{17283} \\
\Id{\detokenize{enwiki-2013}}                                    & \textit{\numprint{2178255}}         & \numprint{13612} & \numprint{2177965}         & \numprint{17336} & \textbf{\numprint{2178436}}  & \numprint{17408} & \numprint{2178327}           & \numprint{17697} \\
\Id{\detokenize{eu-2005}}                                        & \numprint{452296}                   & \numprint{11995} & \numprint{452311}          & \numprint{22968} & \numprint{452342}            & \numprint{5512}  & \textbf{\numprint{452353}}   & \numprint{2332} \\
\Id{\detokenize{hollywood-2011}}                                 & \textbf{\numprint{523402}}          & \numprint{33}    & \textbf{\numprint{523402}} & \numprint{101}   & \textbf{\numprint{523402}}   & \numprint{9}     & \textbf{\numprint{523402}}   & \numprint{17} \\
\Id{\detokenize{libimseti}}                                      & \textit{\numprint{127288}}          & \numprint{8250}  & \numprint{127284}          & \numprint{9308}  & \textbf{\numprint{127292}}   & \numprint{102}   & \textbf{\numprint{127292}}   & \numprint{16747} \\
\Id{\detokenize{ljournal-2008}}                                  & \numprint{2970236}                  & \numprint{428}   & \numprint{2970887}         & \numprint{16571} & \textbf{\numprint{2970937}}  & \numprint{36}    & \textbf{\numprint{2970937}}  & \numprint{41} \\
\Id{\detokenize{orkut}}                                          & \textit{\textbf{\numprint{839073}}} & \numprint{17764} & \numprint{839001}          & \numprint{17933} & \numprint{839004}            & \numprint{19765} & \numprint{806244}            & \numprint{34197} \\
\Id{\detokenize{web-Stanford}}                                   & \textit{\numprint{163384}}          & \numprint{5938}  & \numprint{163382}          & \numprint{10924} & \numprint{163388}            & \numprint{35}    & \textbf{\numprint{163390}}   & \numprint{12} \\
\Id{\detokenize{webbase-2001}}                                   & \numprint{79998332}                 & \numprint{35240} & \numprint{80002845}        & \numprint{35922} & \numprint{80009041}          & \numprint{30960} & \textbf{\numprint{80009820}} & \numprint{31954} \\
\Id{\detokenize{wikilinks}}                                    & \numprint{19404530}                 & \numprint{21069} & \numprint{19416213}        & \numprint{34085} & \numprint{19418693}          & \numprint{23133} & \textbf{\numprint{19418724}} & \numprint{854} \\
\Id{\detokenize{youtube}}                                        & \numprint{857914}                   & $<1$             & \textbf{\numprint{857945}} & \numprint{93}    & \textbf{\numprint{857945}}   & $<1$             & \textbf{\numprint{857945}}   & \numprint{2} \\
%\hline
\multicolumn{9}{l}{\bf Road networks:}\\[0.2em]
%\hline
\Id{\detokenize{europe}}                                            & \numprint{9267573}                  & \numprint{15622} & \numprint{9267587}         & \numprint{28450} & \numprint{9267804}           & \numprint{27039} & \textbf{\numprint{9267809}}  & \numprint{115} \\
\Id{\detokenize{USA-road}}                                       & \numprint{12426557}                 & \numprint{10490} & \numprint{12426582}        & \numprint{31583} & \numprint{12427819}          & \numprint{32490} & \textbf{\numprint{12428099}} & \numprint{4799} \\
%\hline
\multicolumn{9}{l}{\bf Meshes:}\\[0.2em]
%\hline
\Id{\detokenize{buddha}}                                         & \numprint{480795}                   & \numprint{17895} & \textbf{\numprint{480808}} & \numprint{17906} & \numprint{480592}            & \numprint{16695} & \numprint{479905}            & \numprint{17782} \\
\Id{\detokenize{bunny}}                                          & \numprint{32283}                    & \numprint{13258} & \numprint{32287}           & \numprint{13486} & \numprint{32110}             & \numprint{14185} & \textbf{\numprint{32344}}    & \numprint{1309} \\
\Id{\detokenize{dragon}}                                         & \textit{\textbf{\numprint{66501}}}  & \numprint{15203} & \numprint{66496}           & \numprint{14775} & \numprint{66386}             & \numprint{16577} & \numprint{66447}             & \numprint{3456} \\
\Id{\detokenize{feline}}                                         & \textit{\numprint{18846}}           & \numprint{15193} & \numprint{18844}           & \numprint{10547} & \numprint{18732}             & \numprint{15055} & \textbf{\numprint{18851}}    & \numprint{706} \\
\Id{\detokenize{gameguy}}                                        & \numprint{20662}                    & \numprint{6868}  & \numprint{20674}           & \numprint{12119} & \numprint{20655}             & \numprint{7467}  & \textbf{\numprint{20727}}    & \numprint{191} \\
\Id{\detokenize{venus}}                                          & \textbf{\numprint{2684}}            & \numprint{507}   & \textbf{\numprint{2684}}   & \numprint{528}   & \numprint{2664}              & \numprint{9}     & \numprint{2683}              & \numprint{74} \\
\bottomrule
 \end{tabulary}
 \vspace*{-.5cm}
\end{table}

We further test \kermis, which first kernelizes the graph using the advanced reductions from \redumis, removes $1$\% of the highest-degree vertices, and then runs \arw{} on the remaining graph.
On $8$ instances, \kermis{} finds a better solution than \redumis. However, kernelization and cutting take a long time (over 3 hours for \Id{sk-2005}, 10 hours for \Id{uk-2007}), and therefore \kermis{} is much slower to get to a high-quality solution than \online.
Thus, our experiments show that the full set of reductions is not always necessary, especially when the goal is to get a high-quality solution quickly.
This also further justifies our choice of cutting, as the solution quality of \kermis{} remains high.
On the other hand, instances \Id{as-Skitter-big}, \Id{ljournal-2008}, and \Id{youtube} are solved quickly with advanced reduction rules.
\else
Further information about overall solution quality can be found in Appendix~\ref{sec:quality}.
\fi

\vspace*{-.25cm}
\section{Conclusion and Future Work}
\vspace*{-.25cm}
We have shown that applying reductions on the fly during local search leads to high-quality independent sets quickly. Furthermore, cutting few high-degree vertices has little effect on the quality of independent sets found during local search. Lastly, by kernelizing with advanced reduction rules, we can further speed up local search for high-quality independent sets, in the long-run---rivaling the current best heuristic algorithms for complex networks. Determining which reductions give a desirable balance between high-quality results and speed is an interesting topic for future research. While we believe that \online{} gives a nice balance, it is possible that further reductions may achieve higher-quality results even faster.

\bibliographystyle{plain}
\bibliography{phdthesiscs}

\begin{thebibliography}{10}

\bibitem{abu-khzam-2007}
N.~Faisal Abu-Khzam, R.~Michael Fellows, A.~Michael Langston, and Henry~W.
  Suters.
\newblock Crown structures for vertex cover kernelization.
\newblock {\em Theory of Computing Systems}, 41(3):411--430, 2007.

\bibitem{akiba-tcs-2016}
T.~Akiba and Y.~Iwata.
\newblock Branch-and-reduce exponential/fpt algorithms in practice: A case
  study of vertex cover.
\newblock {\em Theoretical Computer Science}, 609, Part 1:211--225, 2016.

\bibitem{AndradeRW12}
D.~V. Andrade, M.~G.~C. Resende, and R.~F. Werneck.
\newblock {Fast Local Search for the Maximum Independent Set Problem}.
\newblock {\em Journal of Heuristics}, 18(4):525--547, 2012.

\bibitem{benchmarksfornetworksanalysis}
D.~A. Bader, H.~Meyerhenke, P.~Sanders, C.~Schulz, A.~Kappes, and D.~Wagner.
\newblock {Benchmarking for Graph Clustering and Partitioning}.
\newblock In {\em Encyclopedia of Social Network Analysis and Mining}, pages
  73--82. Springer, 2014.

\bibitem{batsyn-mcs-ils-2014}
M.~Batsyn, B.~Goldengorin, E.~Maslov, and P.~Pardalos.
\newblock Improvements to mcs algorithm for the maximum clique problem.
\newblock {\em Journal of Combinatorial Optimization}, 27(2):397--416, 2014.

\bibitem{battiti2001reactive}
R.~Battiti and M.~Protasi.
\newblock {Reactive Local Search for the Maximum Clique Problem}.
\newblock {\em Algorithmica}, 29(4):610--637, 2001.

\bibitem{bourgeois-2012}
N.~Bourgeois, B.~Escoffier, V.. Paschos, and J.M. van Rooij.
\newblock Fast algorithms for max independent set.
\newblock {\em Algorithmica}, 62(1-2):382--415, 2012.

\bibitem{butenko-2002}
S.~Butenko, P.~Pardalos, I.~Sergienko, V.~Shylo, and P.~Stetsyuk.
\newblock Finding maximum independent sets in graphs arising from coding
  theory.
\newblock In {\em Proceedings of the 2002 ACM Symposium on Applied Computing},
  SAC '02, pages 542--546, New York, NY, USA, 2002. ACM.

\bibitem{chen-1999}
Jianer Chen, Iyad~A. Kanj, and Weijia Jia.
\newblock Vertex cover: Further observations and further improvements.
\newblock {\em Journal of Algorithms}, 41(2):280--301, 2001.

\bibitem{DSSW09}
D.~Delling, P.~Sanders, D.~Schultes, and D.~Wagner.
\newblock {Engineering Route Planning Algorithms}.
\newblock In {\em Algorithmics of Large and Complex Networks}, volume 5515 of
  {\em LNCS State-of-the-Art Survey}, pages 117--139. Springer, 2009.

\bibitem{demetrescu2009shortest}
C.~Demetrescu, A.~V. Goldberg, and D.~S. Johnson.
\newblock {\em The Shortest Path Problem: 9th DIMACS Implementation Challenge},
  volume~74.
\newblock AMS, 2009.

\bibitem{feo1994greedy}
T.~A. Feo, M.~G.~C. Resende, and S.~H. Smith.
\newblock {A Greedy Randomized Adaptive Search Procedure for Maximum
  Independent Set}.
\newblock {\em Operations Research}, 42(5):860--878, 1994.

\bibitem{fomin-2010}
F.V. Fomin and D.~Kratsch.
\newblock {\em Exact Exponential Algorithms}.
\newblock Springer, 2010.

\bibitem{DBLP:books/fm/GareyJ79}
M.~R. Garey and David~S. Johnson.
\newblock {\em {Computers and Intractability: {A} Guide to the Theory of
  NP-Completeness}}.
\newblock W. H. Freeman, 1979.

\bibitem{gemsa2014dynamiclabel}
A.~Gemsa, M.~Nöllenburg, and I.~Rutter.
\newblock Evaluation of labeling strategies for rotating maps.
\newblock In {\em Experimental Algorithms}, volume 8504 of {\em LNCS}, pages
  235--246. Springer, 2014.

\bibitem{grosso2004combining}
A.~Grosso, M.~Locatelli, and F.~Della~C.
\newblock {Combining Swaps and Node Weights in an Adaptive Greedy Approach for
  the Maximum Clique Problem}.
\newblock {\em Journal of Heuristics}, 10(2):135--152, 2004.

\bibitem{grosso2008simple}
A.~Grosso, M.~Locatelli, and W.~Pullan.
\newblock {Simple Ingredients Leading to Very Efficient Heuristics for the
  Maximum Clique Problem}.
\newblock {\em Journal of Heuristics}, 14(6):587--612, 2008.

\bibitem{hansen2004variable}
P.~Hansen, N.~Mladenovi{\'c}, and D.~Uro{\v{s}}evi{\'c}.
\newblock {Variable Neighborhood Search for the Maximum Clique}.
\newblock {\em Discrete Applied Mathematics}, 145(1):117--125, 2004.

\bibitem{iwata-2014}
Y.~Iwata, K.~Oka, and Y.~Yoshida.
\newblock {Linear-time FPT Algorithms via Network Flow}.
\newblock In {\em Proceedings of 25th ACM-SIAM Symposium on Discrete
  Algorithms}, SODA '14, pages 1749--1761. SIAM, 2014.

\bibitem{katayama2005effective}
K.~Katayama, A.~Hamamoto, and H.~Narihisa.
\newblock {An Effective Local Search for the Maximum Clique Problem}.
\newblock {\em Information Processing Letters}, 95(5):503--511, 2005.

\bibitem{kieritz-contraction-2010}
T.~Kieritz, D.~Luxen, P.~Sanders, and C.~Vetter.
\newblock Distributed time-dependent contraction hierarchies.
\newblock In Paola Festa, editor, {\em Experimental Algorithms}, volume 6049 of
  {\em LNCS}, pages 83--93. Springer Berlin Heidelberg, 2010.

\bibitem{kunegis-2013}
J{\'e}r\^{o}me Kunegis.
\newblock Konect: The koblenz network collection.
\newblock In {\em Proceedings of the 22Nd International Conference on World
  Wide Web}, WWW '13 Companion, pages 1343--1350, Republic and Canton of
  Geneva, Switzerland, 2013. International World Wide Web Conferences Steering
  Committee.

\bibitem{webgraphWS}
University of~Milano Laboratory~of Web~Algorithms.
\newblock Datasets.

\bibitem{lammSEA2015}
S.~Lamm, P.~Sanders, and C.~Schulz.
\newblock {Graph Partitioning for Independent Sets}.
\newblock In {\em Proceedings of the 14th International Symposium on
  Experimental Algorithms (SEA'15)}, volume 8504, pages 68--81. Springer, 2015.

\bibitem{kamis-alenex-2016}
S.~Lamm, P.~Sanders, C.~Schulz, D.~Strash, and R.~F. Werneck.
\newblock {\em Finding Near-Optimal Independent Sets at Scale}, chapter~11,
  pages 138--150.
\newblock 2016.

\bibitem{external-memory-ils}
Y.~Liu, J.~Lu, H.~Yang, X.~Xiao, and Z.~Wei.
\newblock Towards maximum independent sets on massive graphs.
\newblock {\em Proc. VLDB Endow.}, 8(13):2122--2133, September 2015.

\bibitem{nemhauser-1975}
G.L. Nemhauser and Jr. Trotter, L.E.
\newblock Vertex packings: Structural properties and algorithms.
\newblock {\em Mathematical Programming}, 8(1):232--248, 1975.

\bibitem{pullan2006dynamic}
W.~J. Pullan and H.~H. Hoos.
\newblock {Dynamic Local Search for the Maximum Clique Problem}.
\newblock {\em Journal of Artifical Intelligence Research(JAIR)}, 25:159--185,
  2006.

\bibitem{segundo-recoloring}
P.~San~Segundo, F.~Matia, D.~Rodriguez-Losada, and M.~Hernando.
\newblock An improved bit parallel exact maximum clique algorithm.
\newblock {\em Optimization Letters}, 7(3):467--479, 2013.

\bibitem{segundo-bitboard-2011}
P.~San~Segundo, D.~Rodríguez-Losada, and Agustín J.
\newblock An exact bit-parallel algorithm for the maximum clique problem.
\newblock {\em Computers \& Operations Research}, 38(2):571--581, 2011.

\bibitem{sander-mesh-2008}
P.~V. Sander, D.~Nehab, E.~Chlamtac, and H.~Hoppe.
\newblock Efficient traversal of mesh edges using adjacency primitives.
\newblock {\em ACM Trans. Graph.}, 27(5):144:1--144:9, December 2008.

\bibitem{tarjan-1977}
R.~E. Tarjan and A.~E. Trojanowski.
\newblock Finding a maximum independent set.
\newblock {\em SIAM Journal on Computing}, 6(3):537--546, 1977.

\bibitem{tomita-recoloring}
E.~Tomita, Y.~Sutani, T.~Higashi, S.~Takahashi, and M.~Wakatsuki.
\newblock A simple and faster branch-and-bound algorithm for finding a maximum
  clique.
\newblock In Md.~Saidur Rahman and Satoshi Fujita, editors, {\em WALCOM:
  Algorithms and Computation}, volume 5942 of {\em LNCS}, pages 191--203.
  Springer Berlin Heidelberg, 2010.

\bibitem{xiang-2013}
J.~Xiang, C.~Guo, and A.~Aboulnaga.
\newblock Scalable maximum clique computation using mapreduce.
\newblock In {\em Data Engineering (ICDE), 2013 IEEE 29th International
  Conference on}, pages 74--85, April 2013.

\bibitem{Xiao201392}
M.~Xiao and H.~Nagamochi.
\newblock Confining sets and avoiding bottleneck cases: A simple maximum
  independent set algorithm in degree-3 graphs.
\newblock {\em Theoretical Computer Science}, 469:92 -- 104, 2013.

\end{thebibliography}

\newpage
\appendix

\section{Reductions}
\label{sec:reductions}
We briefly describe the reduction rules from Akiba and Iwata~\cite{akiba-tcs-2016}. Each of these exact reductions allow us to choose vertices that are in some MIS by following simple rules. If an MIS is found on the kernel graph $\mathcal{K}$, then each reduction may be undone, producing an MIS in the original graph. Refer to Akiba and Iwata~\cite{akiba-tcs-2016} for a more thorough discussion, including implementation details.

\noindent\emph{Pendant vertices:} Any vertex $v$ of degree one, called a \emph{pendant}, is in some MIS; therefore $v$ and its neighbor $u$ can be removed from $G$.

\noindent\emph{Vertex folding~\cite{chen-1999}:} For a vertex $v$ with degree 2 whose neighbors $u$ and $w$ are not adjacent, either $v$ is in some MIS, or both $u$ and $w$ are in some MIS. Therefore, we can contract $u$, $v$, and $w$ to a single vertex $v'$ and decide which vertices are in the MIS later.

\ifFull
\noindent\emph{Linear Programming:}
First introduced by Nemhauser and Trotter~\cite{nemhauser-1975} for the vertex packing problem, they present a linear programming relaxation with a half-integral solution (i.e., using only values 0, 1/2, and 1) which can be solved using bipartite matching. Their relaxation may be formulated for the independent set problem as follows: maximize $\sum_{v\in V}{x_v}$, such at for each edge $(u, v) \in E$, $x_u + x_v \leq 1$ and for each vertex $v \in V$, $x_v \geq 0$. Vertices with value 1 must be in the MIS, and therefore are added to the solution. We use the further improvement from Iwata, Oka, and Yoshida~\cite{iwata-2014}, which computes a solution whose half-integral part is minimal.
\else
\noindent\emph{Linear Programming:}
A well-known~\cite{nemhauser-1975} linear programming relaxation for the MIS problem with a half-integral solution (i.e., using only values 0, 1/2, and 1) can be solved using bipartite matching: maximize $\sum_{v\in V}{x_v}$ such that $\forall (u, v) \in E$, $x_u + x_v \leq 1$ and $\forall v \in V$, $x_v \geq 0$. Vertices with value 1 must be in the MIS and can thus be removed from $G$ along with their neighbors. We use an improved version~\cite{iwata-2014} that computes a solution whose half-integral part is minimal. % the further improvement from Iwata, Oka, and Yoshida
\fi

\noindent\emph{Unconfined~\cite{Xiao201392}:} Though there are several definitions of \emph{unconfined} vertex in the literature, we use the simple one from Akiba and Iwata~\cite{akiba-tcs-2016}. A vertex $v$ is \emph{unconfined} when determined by the following simple algorithm. First, initialize $S = \{v\}$. Then find a $u \in N(S)$ such that $|N(u) \cap S| = 1$ and $|N(u) \setminus N[S]|$ is minimized. If there is no such vertex, then $v$ is confined. If $N(u) \setminus N[S] = \emptyset$, then $v$ is unconfined.  If $N(u)\setminus N[S]$ is a single vertex $w$, then add $w$ to $S$ and repeat the algorithm. Otherwise, $v$ is confined. Unconfined vertices can be removed from the graph, since there always exists an MIS $\mathcal{I}$ that contains no unconfined vertices.

\ifFull
\noindent\emph{Twin:} This is a generalization of the vertex folding rule. Suppose there are two vertices $u$ and $v$ that have degree 3 and share the same neighborhood. If $u$'s neighborhood $N(u)$ induces a graph with edges, then $u$ and $v$ are added to the independent set and $u$, $v$, and their neighborhoods are removed from the graph. Otherwise, vertices in $N(u)$ may belong in the independent set. We still remove $u$, $v$, and their neighborhoods, and add a new gadget vertex $w$ to the graph with edges to $u$'s two-neighborhood (vertices at a distance 2 from $u$). If $w$ is in some MIS, none of $u$'s two-neighbors are in the independent set, and therefore $N(u)$ is part of the independent set. Otherwise, if $w$ is not in the MIS, then some of $u$'s two-neighbors are in the independent set, and therefore $u$ and $v$ are added to the independent set. Thus, the twin reduction adds an additional two vertices to the computed independent set.
\else
\noindent\emph{Twin~\cite{Xiao201392}:} Let $u$ and $v$ be vertices of degree 3 with $N(u) = N(v)$. If $G[N(u)]$ has edges, then add $u$ and $v$ to $\mathcal{I}$ and remove $u$, $v$, $N(u)$, $N(v)$ from $G$. Otherwise, some vertices in $N(u)$ may belong to some MIS $\mathcal{I}$. We still remove $u$, $v$, $N(u)$ and $N(v)$ from $G$, and add a new gadget vertex $w$ to $G$ with edges to $u$'s two-neighborhood (vertices at a distance 2 from $u$). If $w$ is in the computed MIS, then none of $u$'s two-neighbors are $\mathcal{I}$, and therefore $N(u) \subseteq \mathcal{I}$. Otherwise, if $w$ is not in the computed MIS, then some of $u$'s two-neighbors are in $\mathcal{I}$, and therefore $u$ and $v$ are added to $\mathcal{I}$.
\fi

\noindent\emph{Alternative:} Two sets of vertices $A$ and $B$ are set to be \emph{alternatives} if $|A| = |B| \geq 1$ and there exists an MIS $\mathcal{I}$ such that $\mathcal{I}\cap(A\cup B)$ is either $A$ or $B$. Then we remove $A$ and $B$ and $C = N(A)\cap N(B)$ from $G$ and add edges from each $a \in N(A)\setminus C$ to each $b\in N(B)\setminus C$.
Then we add either $A$ or $B$ to $\mathcal{I}$, depending on which neighborhood has vertices in $\mathcal{I}$. Two structures are detected as alternatives. First, if $N(v)\setminus \{u\}$ induces a complete graph, then $\{u\}$ and $\{v\}$ are alternatives (a \emph{funnel}). Next, if there is a cordless 4-cycle $a_1b_1a_2b_2$ where each vertex has at least degree 3. Then sets $A=\{a_1, a_2\}$ and $B=\{b_1, b_2\}$ are alternatives when $|N(A) \setminus B| \leq 2$, $|N(A) \setminus B| \leq 2$, and $N(A) \cap N(B) = \emptyset$.

\noindent\emph{Packing~\cite{akiba-tcs-2016}:} %A full description of the packing reduction is beyond the scope of this article. However, we briefly describe the intuition behind the reduction. 
Given a non-empty set of vertices $S$, we may specify a \emph{packing constraint} $\sum_{v\in S}x_v \leq k$, where $x_v$ is 0 when $v$ is in some MIS $\mathcal{I}$ and 1 otherwise. Whenever a vertex $v$ is excluded from $\mathcal{I}$ (i.e., in the unconfined reduction), we remove $x_v$ from the packing constraint and decrease the upper bound of the constraint by one. Initially, packing constraints are created whenever a vertex $v$ is excluded or included into the MIS. The simplest case for the packing reduction is when $k$ is zero: all vertices must be in $\mathcal{I}$ to satisfy the constraint. Thus, if there is no edge in $G[S]$, $S$ may be added to $\mathcal{I}$, and $S$ and $N(S)$ are removed from $G$. Other cases are much more complex. Whenever packing reductions are applied, existing packing constraints are updated and new ones are added.

\section{Overall Solution Quality}
\label{sec:quality}
Next, we show that \online{} has high solution quality when given a time limit for searching ($5$ hours for normal graphs, $10$ hours for huge graphs).
Table~\ref{table:avg} shows the average solution size over the three runs.
Although long-run quality is not the goal of the \online{} algorithm, in $11$ instances \online{} finds a larger independent set than \arw, and in $4$ instances \online{} finds the largest solution in the time limit.
As seen in Table~\ref{table:avg}, \online{} also finds a solution within $0.1$\% of the best solution found by any algorithm for all graphs.
However, in general \online{} finds lower-quality solutions than \redumis, which we believe is from high-degree
cutting removing vertices in large independent sets.
Nonetheless, as this shows, even when cutting out $1$\% of the vertices, the solution quality remains high.

We further test \kermis, which first kernelizes the graph using the advanced reductions from \redumis, removes $1$\% of the highest-degree vertices, and then runs \arw{} on the remaining graph.
On $8$ instances, \kermis{} finds a better solution than \redumis. However, kernelization and cutting take a long time (over 3 hours for \Id{sk-2005}, 10 hours for \Id{uk-2007}), and therefore \kermis{} is much slower to get to a high-quality solution than \online.
Thus, our experiments show that the full set of reductions is not always necessary, especially when the goal is to get a high-quality solution quickly.
This also further justifies our choice of cutting, as the solution quality of \kermis{} remains high.
On the other hand, instances \Id{as-Skitter-big}, \Id{ljournal-2008}, and \Id{youtube} are solved quickly with advanced reduction rules.

\end{document}